# Ultrasonic Nanofabrication with an AFM

Ultrasound Facilitates Nanolithography and Nanomanipulation

Ultrasonic AFM may improve fabrication technologies on the nanometer scale. In the presence of ultrasonic vibration, hard surfaces can be indented and scratched with the tip of a soft cantilever, due to its inertia. Ultrasound reduces or even eliminates friction, and hence modifies the tip-nanoparticle-surface interactions in AFM manipulation. The subsurface sensitivity of the technique makes feasible the purposed manipulation of subsurface nanoscale features by ultrasonic actuation.

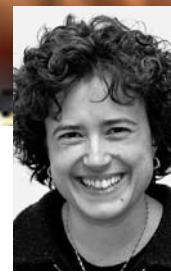

M. Teresa Cuberes

## Ultrasonic Atomic Force Microscopies

Information of ultrasonic vibration on the nanoscale has recently become accessible by a new family of Scanning Probe Microscopy techniques based on the use of Atomic Force Microscopy (AFM) with ultrasound excitation [1]. Among them, the techniques of Ultrasonic Force Microscopy (UFM) [2] and Heterodyne Force Microscopy (HFM) [3] rely in the so-called „mechanical-diode" effect [4], in which a cantilever tip is in contact with the sample surface and normal ultrasonic vibration is excited at the tip-sample contact (see fig. 1). If the excitation frequency is high enough, or is not coincident with a high-order cantilever contact resonance, the cantilever will not be able to linearly follow the surface vibration due to its inertia. Nevertheless, if the ultrasonic excitation amplitude is sufficiently high that the tip-sample distance varies over the nonlinear tip-sample force interaction regime, the cantilever experiences a static force during the time that the ultrasonic excitation is acting. This force is the so-called "ultrasonic force", and can be understood as the net force that acts upon the cantilever during a complete ultrasonic cycle, due to the nonlinearity of the tip-sample interaction force. The cantilever behaves then as a mechanical diode and deflects when the tip-sample contact vibrates at ultrasonic frequencies of sufficiently high amplitude. The magnitude of the ultrasonic force or of the ultrasonic-force-induced additional cantilever deflection (UFM signal) is dependent on the details of the tip-sample interaction force, and hence on material properties such as elasticity and adhesion. Therefore, UFM allow us to discern nanoscale topographic regions with distinct elastic contrast, as shown in figure 2, where images of Sb nanoparticles on Highly Oriented Pyrolitic Graphite (HOPG) are displayed. Remarkably, in this case, the UFM contrast reveals stiffness variations even within individual Sb particles [5]. Recently, a novel ultrasonic AFM mode, namely Mechanical-Diode Ultrasonic Friction Force Microscopy (MD-UFFM) based on the lateral mechanical diode effect has additionally been proposed for the study of friction and lubrication on the nanoscale, in the presence of surface shear ultrasonic vibration [6].

## Ultrasonic Nanofabrication

Ultrasonic AFM techniques provide a means to monitor ultrasonic vibration at the nanoscale, and open up novel opportunities to improve nanofabrication technologies [7].

In the presence of ultrasonic vibration, the tip of a soft cantilever can dynamically indent hard samples due to its inertia. In addition, it has been demonstrated that ultrasound reduces or even eliminates nanoscale friction [8]. Typical top-down approaches that rely in the AFM are based on the use of a cantilever tip that acts as a plow or as an engraving tool. The ability of the AFM tip to respond inertially to ultrasonic vibration excited perpendicular to the sample surface and dynamically indent hard samples may facilitate the nanoscale machining of semiconductors or engineering ceramics in a reduced time. Figure 3 demonstrates the machining of nanotrenches and holes on a silicon sample in the presence of ultrasonic vibration. Interestingly, no debris is found in the proximity of lithographed areas. Figure 3 (a) refer to results performed using a cantilever with nominal stiffness comprised between 28–91 $Nm^{-1}$ and a diamond-coated tip. Figure 4 (b) refer to results achieved using a cantilever with nominal stiffness 0.11 $Nm^{-1}$ and a SiN tip; in the absence of ultrasound, it was not possible to scratch the Si surface using such a soft cantilever. In the machining of soft materials, as for instance plastic coatings, the ultrasonic-induced reduction of nanoscale friction may permit eventual finer features and improved surface quality in quasi-static approaches. In [9], an in-plane acoustic wave coupled to the sample support was used to enhance the intermittent force exerted by the tip in dynamic AFM nanomachining of thin polymer resist films.

In bottom-up approaches, ultrasound may assist in the self-assembly or AFM manipulation of nanostructures [7]. Effects such as sonolubrication and acoustic levitation have been studied at the microscale. These phenomena may facilitate a tip-induced motion of nano-objects. In the manipulation of nanoparticles (NPs) on surfaces with the tip of an AFM cantilever, when ultrasound is excited at a sample surface both tip-particle and particle-surface frictional properties change [10]. Moreover, the excitation of NP high-frequency internal vibration modes may also modify the NP dynamic response, and introduce novel mechanisms of particle motion. Some of the opportunities in ultrasonic-assisted AFM manipulation are illustrated in figure 4, which show images of Au NPs on a silicon surface. The surface was covered by poly-l-lysine to prevent that the Au NPs were swept away by the AFM tip. Nevertheless, when scanning in contact mode in the absence of ultrasound, most NPs were inevitably swept away by the tip. Scanning in the same conditions that led to a NP displacement but in the presence of surface ultrasonic vibration, with an appropriate election of the ultrasonic amplitude, the NPs remained undisturbed. In figure 4, two NP were displaced while recording the images, which have been pointed out by blue arrows. A close inspection of the data in figure 4 reveals traces of the UFM and LFM responses from those moving particles while being in motion (see areas enclosed by ellipses and rectangles). The study of the UFM response of a moving NP may allow us to learn about the dynamic mechanisms of NP displacement across surfaces. Consistently with the fact that ultrasound eliminates friction, no frictional contrast is distinguished on the undisturbed Au NPs in the LFM images. However, from the comparison of the traces of the two moving NPs in the forward and backward LFM scans, friction during the tip-induced NP motion is apparent. Controlled and accurate measurements of lateral and ultrasonic forces exerted by individual NPs when in motion under tip ultrasonic actuation may bring about a wealth of information about the dissipated energy, ultrasonic lubrication effects, NP dynamics, etc.

Eventually, it should be pointed out that the sensitivity of ultrasonic-AFM to subsurface features makes feasible to monitor subsurface

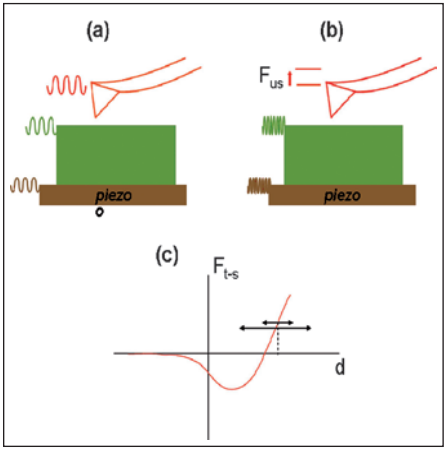

**Fig. 1:** Detection of surface vibration with the tip of an AFM cantilever. (a) At low frequencies, the tip follows the surface vibration. (b) In the high-frequency regime, for sufficiently high vibration amplitudes, tip experiences an ultrasonic force $F_{us}$. (c) Tip-sample force $F_{t-s}$ versus tip-sample distance d curve.

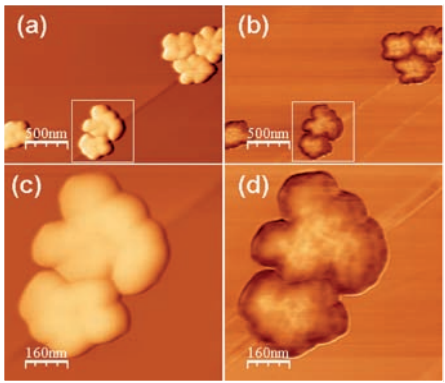

**Fig. 2:** Sb NPs on HOPG. (a, b) Simultaneously recorded contact-mode AFM topography (a) and UFM image (b). (c, d) Simultaneously recorded high-resolution images from the areas enclosed by white squares in (a), (b). Set point: 1 nN; Kc: 0.11 Nm$^{-1}$; UFM parameters: 2.2 MHz, 8 Vpp. Contrast in UFM indicates stiffness variations.

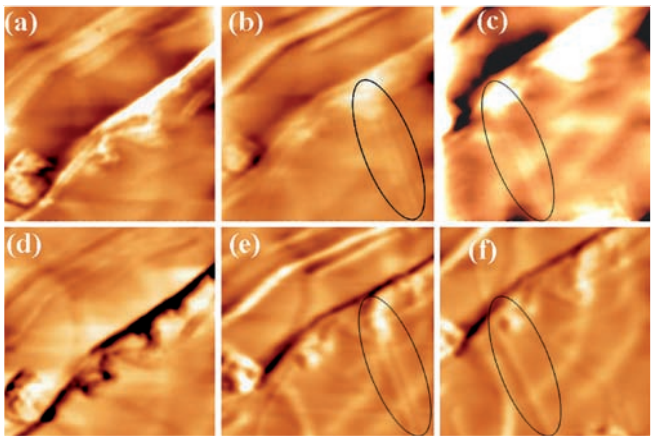

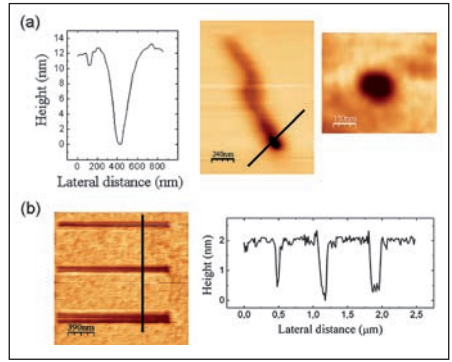

**Fig. 3:** AFM nanomachining of trenches and holes on Si(111) in the presence of normal surface ultrasonic vibration of ~5 MHz. (a) Scratch and indentation with a diamond-coated cantilever tip. Rt: 35 nm. Kc: 28-91 Nm$^{-1}$. (b) Nanotrenches formed in 50, 75 and 100 cycles respectively, at a load of ~40 nN, with a pyramidal SiN cantilever tip. Kc: 0.11 Nm$^{-1}$.

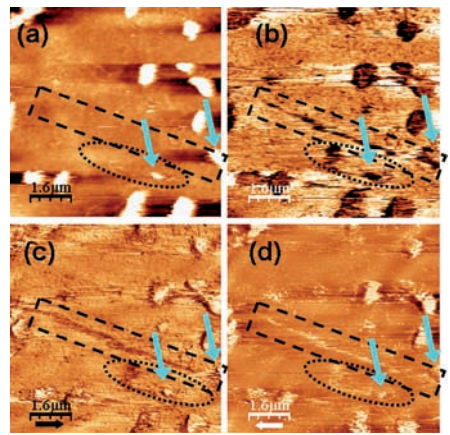

**Fig. 4:** Displacement of Au NP induced by the tip of an AFM cantilever in the presence of normal surface ultrasonic vibration of ~ 2.6 MHz. (a-d) were simultaneously recorded. (a) Contact-mode AFM topography. (b) UFM; (c) LFM forward scan; (d) LFM backward scan

modifications [7]. We have recently demonstrated that actuation with an AFM tip in the presence of ultrasonic vibration can produce stacking changes of extended grapheme layers, and induce permanent displacements of buried dislocations in Highly Oriented Pyrolytic Graphite (HOPG). This effect is illustrated in figure 5. In the presence of normal surface ultrasonic vibration, both AFM and LFM images reveal subsurface features [1]. Subsurface modification was brought about in this case by scanning in contact mode, with high set-point forces, and high surface ultrasonic excitation amplitudes [7].

**Fig. 5:** Lateral displacement of a subsurface dislocation in HOPG by ultrasonic tip actuation (a) Topography of the HOPG surface in AFM contact mode: (700 x 700) nm; Fo=105 nN. (b,c) Ultrasonic-AFM images recorded in sequence over nearly the same surface region: a=2.15 MHz (b) A=3.5 Vpp (c) A=2.4 Vpp. (d–f) LFM forward scan, simultaneously recorded with (a–b) respectively, with the same parameters.

## Summary

Ultrasonic AFM techniques provide a means to monitor ultrasonic vibration at the nanoscale, and open up novel opportunities in nanofabrication technologies. The use of ultrasound may improve both down-top and bottom-down approaches in nanofabrication, facilitating the patterning of nanoscale surface features, the manipulation or self-assembly of nanostructures, and possibly the controlled subsurface manipulation of buried nano-objects.

## Acknowledgments

The author thanks J. J. Martinez and A. Lusvardi for assistance in the UFM lab. The samples of Sb NPs were provided by C. Ritter and U. Schwarz. The Au NPs were provided by M. A. Gonzalez and M.P. Morales. Financial support from the JCCM (Junta de Comunidades de Castilla-La Mancha) under project PBI-05-018 is gratefully acknowledged.

**Contact:**
**Dr. M. Teresa Cuberes**
Engineering School Professor
University of Castilla-La Mancha
Applied Mechanics and Project Engineering
Laboratory of Nanotechnology
Almadén, Spain
Tel.: +34 902 204100 ext. 6045
Fax: +34 926 264401
teresa.cuberes@uclm.es
www.uclm.es